\documentclass[aps,prl,preprint,showpacs,groupedaddress]{revtex4}
\begin{document}
\title{THE ENERGY DISTRIBUTION OF ATOMS IN THE FIELD OF THERMAL RADIATION}
\author{Fedor V.Prigara}
\affiliation{Institute of Microelectronics and Informatics,
Russian Academy of Sciences, 21 Universitetskaya, Yaroslavl
150007, Russia} \email{fprigara@imras.yar.ru}
\date{\today}

\begin{abstract}
 Using the principle of detailed balance and the assumption on the absorption
cross-section consistent with available astrophysical data, we
obtain the energy distribution of atoms in the field of thermal
blackbody radiation and show that this distribution diverges from
the Boltzmann law. There is an inversion of the high energy level
population at sufficiently high temperatures.
\end{abstract}
\maketitle

In the unified model of compact radio sources, radio emission from
active galactic nuclei and pulsars is treated as thermal radiation
from an accretion disk amplified by a maser mechanism [1]. A maser
amplification of thermal radio emission in continuum produces the
high brightness temperatures of compact radio sources and a rapid
variability of total and polarized flux density, that is
characteristic for non-saturated maser sources. In particular,
pulsars signals show a variability on every observable time scale
up to nanoseconds [2,3]. The brightness temperatures of OH masers
have the magnitude $T_{b} < 10^{12}K$, and those of water masers
have the magnitude $T_{b} < 10^{15}K$ [4]. Compact extragalactic
sources (active galactic nuclei) exhibit brightness temperatures
in the range of $10^{10}K$ to $10^{12}K$ [5,6], so these
temperatures have an order of magnitude of those of OH masers.

Maser amplification in continuum is closely connected with the
stimulated origin of thermal radio emission. The induced origin of
thermal radio emission follows from the relations between
Einstein's coefficients for spontaneous and induced emission of
radiation. However, the detailed mechanism of maser amplification
has been unknown so far. In this paper we show that there is an
inversion of the high energy level population at sufficiently high
temperatures which can give rise to the maser amplification in
continuum in a hot plasma.

The original Einstein theory of spontaneous and induced emission
of radiation did not take into account a line width [7]. Here we
show, in particular, that the account for a natural line width
essentially changes the conclusions on the wavelength dependence
of the absorption and stimulated emission cross-sections as well
as the relations between Einstein's coefficients.

Consider two level system (atom) with energy levels $E_{1} $ and
$E_{2} > E_{1} $ which is in equilibrium with thermal blackbody
radiation. We denote as $\nu _{s} = A_{21} $ the number of
transitions from the upper energy level to the lower one per unit
time caused by spontaneous emission of radiation with the
frequency $\omega = \left( {E_{2} - E_{1}}  \right)/\hbar $, where
$\hbar $ is the Planck constant. The number of transitions from
the energy level $E_{2} $ to the energy level $E_{1} $ per unit
time caused by stimulated emission of radiation may be written in
the form

\begin{equation}
\label{eq1}
\nu _{i} = B_{21} B_{\nu}  \Delta \nu
 = \sigma _{21} \frac{{\pi B_{\nu} } }{{\hbar \omega} }\Delta \nu ,
\end{equation}

\noindent
where

\begin{equation}
\label{eq2}
B_{\nu}  = \hbar \omega ^{3}/\left( {2\pi ^{2}c^{2}} \right)\left(
{exp\left( {\hbar \omega /kT} \right) - 1} \right)
\end{equation}

\noindent
is a blackbody emissivity (Planck's function), $\sigma _{21} $ is the
stimulated emission cross-section,$T$ is the temperature, \textit{k} is the
Boltzmann constant, \textit{c} is the speed of light, and $\Delta \nu $ is
the line width.

The number of transitions from the lower level to the upper one per unit
time may be written in the form

\begin{equation}
\label{eq3}
\nu _{12} = B_{12} B_{\nu}  \Delta \nu = \sigma _{12} \frac{{\pi B_{\nu}
}}{{\hbar \omega} }\Delta \nu .
\end{equation}

\noindent
where $\sigma _{12} $ is the absorption cross-section.

Here \textit{A}$_{21}$\textit{,}$B_{21} $ and $B_{12} $ are the
coefficients introduced by Einstein [7], the coefficients $B_{12}
$ and $B_{21} $ being modified to account for the line width
$\Delta \nu $.

We denote as $N_{1} $ and $N_{2} $ the number of atoms occupying the energy
levels $E_{1} $ and $E_{2} $, respectively. The levels $E_{1} $ and $E_{2} $
are suggested for simplicity to be non-degenerated.

In the equilibrium state the full number of transitions from the lower level
to the upper one is equal to the number of reverse transitions :

\begin{equation}
\label{eq4}
N_{1} \nu _{12} = N_{1} B_{12} B_{\nu}  \Delta \nu = N_{2} \nu _{21} = N_{2}
\left( {A_{21} + B_{21} B_{\nu}  \Delta \nu}  \right).
\end{equation}

The line width $\Delta \nu $ is suggested to be equal to the
natural line width [8,9]:

\begin{equation}
\label{eq5}
\Delta \nu = A_{21} + B_{21} B_{\nu}  \Delta \nu .
\end{equation}

It is assumed that the lower level has a zero energy width, i.e. this level
is a ground state.

It follows from the last equation that

\begin{equation}
\label{eq6}
\Delta \nu = A_{21} /\left( {1 - B_{21} B_{\nu} }  \right).
\end{equation}

Substituting this expression in the equation \ref{eq4} , we obtain

\begin{equation}
\label{eq7}
N_{2} /N_{1} = B_{12} B_{\nu}  = \sigma _{12} \frac{{\pi B_{\nu} } }{{\hbar
\omega} }.
\end{equation}

In the limit of high temperatures $T \to \infty $, corresponding to the
range of frequencies $\hbar \omega < kT$, the function $B_{\nu}  \left( {T}
\right)$ is given by the Rayleigh-Jeans formula

\begin{equation}
\label{eq8}
B_{\nu}  = 2kT\nu ^{2}/c^{2},
\end{equation}

\noindent
where $\nu $ is the frequency of radiation , $\nu = \omega /2\pi $.

Since $B_{\nu}  \to \infty $ when $T \to \infty $ , it follows from the
equation (\ref{eq6}) that the coefficient $B_{21} $ is depending on the temperature
\textit{T} in such a manner that $B_{21} B_{\nu}  < 1$ . It is clear that
$B_{21} B_{\nu}  \to 1$ when $T \to \infty $ . The ratio of frequencies of
transitions caused by spontaneous and induced emission of radiation is given
by the expression

\begin{equation}
\label{eq9}
\nu _{s} /\nu _{i} = A_{21} /\left( {B_{21} B_{\nu}  \Delta \nu}  \right) =
\left( {1 - B_{21} B_{\nu} }  \right)/\left( {B_{21} B_{\nu} }  \right)
\quad .
\end{equation}

This ratio is approaching zero when$T \to \infty $ , since $B_{21}
B_{\nu} \to 1$. It means that in the range of frequencies $\hbar
\omega < kT$ thermal radiation is produced by the stimulated
emission, whereas the contribution of the spontaneous emission may
be neglected .

It follows from equations (\ref{eq7}) and (\ref{eq8}) that in the
Rayleigh-Jeans range of frequencies the ratio $N_{2} /N_{1} $ is
given by the formula

\begin{equation}
\label{eq10}
N_{2} /N_{1} = \sigma _{12} \omega kT/\left( {2\pi \hbar c^{2}} \right).
\end{equation}

An analysis of observational data on thermal radio emission from
various astrophysical objects suggests that the absorption
cross-section does not depend on a wavelength of radiation and has
an order of magnitude of an atomic cross-section, $\sigma _{12}
\approx 10^{ - 15}cm^{2}$. Only this assumption is consistent with
the observed spectra of thermal radio emission from major planets,
the spectral indices of radio emission from galactic and
extragalactic sources, and the wavelength dependence of radio
source size.

If the absorption cross-section is constant then the energy distribution of
atoms has a form

\begin{equation}
\label{eq11}
N_{2} /N_{1} = const\left( {E_{2} - E_{1}}  \right)kT.
\end{equation}

For all real values of the temperature and wavelength of radiation
the ratio (\ref{eq11}) is much smaller than unit. For example, at
$\lambda = 1m$ the ratio (\ref{eq11}) is less than 1, if
$T<10^{18}K$. It implies that $N_{1} \approx N$, where \textit{N}
is the total number of atoms. There are no reasons to expect that
the Boltzmann distribution is still valid for an ensemble of atoms
interacting with thermal radiation.

From the relation $B_{21} B_{\nu}  \approx 1$ we obtain the
stimulated emission cross-section in the form

\begin{equation}
\label{eq12}
\sigma _{21} \approx \hbar \omega /\left( {\pi B_{\nu} }  \right) = \lambda
l_{T} /\left( {2\pi}  \right),
\end{equation}

\noindent
where $\lambda $ is the wavelength, and $l_{T} = 2\pi \hbar c/\left( {kT}
\right)$. Thus, Einstein's relation $B_{12} = B_{21} $, equivalent to
$\sigma _{12} = \sigma _{21} $, is not valid in the Rayleigh-Jeans region.

Consider now the Wien region $\hbar \omega > kT$. There are strong
theoretical and observational indications of the stimulated
character of thermal blackbody radiation in the whole range of
spectrum. First, the spectral energy density of thermal blackbody
radiation is described by a single Planck's function at all
frequencies. Second, there are clear observational indications of
the existence of thermal harmonics in stellar spectra and of laser
type sources. The latter are connected with the induced origin of
thermal radiation, similarly to well known maser sources. In
particular, a possible non-saturated X-ray laser source emitting
in the Fe $K_{\alpha}  $line at 6.49 keV was recently discovered
in the radio-loud quasar MG J0414+0534 [10].

Assuming the stimulated origin of thermal blackbody radiation,
i.e. $\nu _{s} < \nu _{i} $ , from equation (\ref{eq8}) one can
obtain the relation $B_{21} B_{\nu} = \sigma _{21} \pi B_{\nu}
/\left( {\hbar \omega}  \right) \approx 1$, and then the
stimulated emission cross-section in the form

\begin{equation}
\label{eq13}
\sigma _{21} \approx \left( {\lambda ^{2}/2\pi}  \right)exp\left( {\hbar
\omega /kT} \right).
\end{equation}

The absorption cross-section is likely to be constant in the whole
range of spectrum, so the ratio $N_{2} /N_{1} $, according to
equation (\ref{eq6}), is given by the formula

\begin{equation}
\label{eq14}
N_{2} /N_{1} \approx \left( {\sigma _{12} \omega ^{2}/2\pi c^{2}}
\right)exp\left( { - \hbar \omega /kT} \right),
\end{equation}

\noindent
which is somewhat similar to the Boltzmann law.

The exact formula for the ratio $N_{2} /N_{1} $ can be obtained
from (\ref{eq6}) :

\begin{equation}
\label{eq15}
N_{2} /N_{1} = \sigma _{12} \omega ^{2}/\left( {2\pi c^{2}} \right)\left(
{exp\left( {\hbar \omega /kT} \right) - 1} \right).
\end{equation}

The function (\ref{eq15}) has a maximum at $\hbar \omega = 1.6kT$.
It means that, in the field of thermal blackbody radiation, the
excited levels with $E_{2} - E_{1} \approx kT$ are the most
populated. Numerically, $\left( {N_{2} /N_{1} } \right)_{max} =
0.8 \times 10^{ - 15}T^{2}$, so the ratio $N_{2} /N_{1} $ is less
than 1, if only $T<3\times10^{7}K$. For the temperatures
$T>3\times10^{7}K$ the population of the levels corresponding to
the maximum of the function (\ref{eq15}) is inverse. It suggests
that laser type sources are more easily realized in X-ray and
gamma-ray range of spectrum than in the optical region. The well
known example is a possible gamma-laser in the Galactic Center
emitting in the line 0.511 MeV [11].

The author is grateful to A.V.Postnikov for useful discussions.

--------------------------------------------------------------------------------------------------------------

[1] F.V.Prigara. Astron. Nachr., Vol. \textbf{324}, No. S1, p.425
(2003).

[2] R.T.Edwards, and B.W.Stappers. Astron. Astrophys. (submitted),
astro-ph/0305266.

[3] M.Vivekanand. Month. Not. Roy. Astron. Soc., \textbf{326}, L33
(2001).

[4] N.G.Bochkarev. \textit{Basic Physics of Interstellar Medium}
(Moscow University Press, Moscow, 1992).

[5] G.C.Bower, and D.C.Backer. Astrophys. J., \textbf{507}, L117
(1998).

[6] K.I.Kellermann, R.C.Vermeulen, J.A.Zensus, and M.H.Cohen.
Astron. J., \textbf{115}, 1295 (1998).

[7] J.R.Singer, \textit{Masers} (Wiley, New York, 1959).

[8] L.D.Landau, and E.M.Lifshitz, \textit{Quantum Mechanics,
Non-Relativistic Theory} (Addison-Wesley, Reading, MA, 1958).

[9] V.B.Berestetskii, E.M.Lifshitz, and L.P.Pitaevskii,
\textit{Quantum Electrodynamics} (Nauka, Moscow, 1989).

[10] G.Chartas, E.Agol, M.Eracleous, G.Garmir, M.W.Bautz, and
N.Moran. Astrophys. J., \textbf{568}, 509 (2002).

[11] R.E.Lingenfelter, and R.Ramaty, in \textit{AIP Proc. 83: The
Galactic Center}, edited by G.Riegler, and R.Blandford (AIP, New
York, 1982).

\end{document}